\documentclass[aps,prb,amsmath,twocolumn,showpacs,superscriptaddress,10pt]{revtex4-1}
\pdfoutput=1
\usepackage{graphicx} 

\newcommand{\eff}{\text{eff}}
\newcommand{\AFM}{\text{AFM}}

\newcommand{\CW}{\text{CW}}
\newcommand{\Dv}{\mathbf D}
\newcommand{\Sv}{\mathbf S}
\newcommand{\Hv}{\mathbf H}
\newcommand{\Lv}{\mathbf L}
\newcommand{\gv}{\mathbf g}

\newcommand{\bv}{\mathbf b}
\newcommand{\cv}{\mathbf c}
\newcommand{\xv}{\mathbf x}
\newcommand{\yv}{\mathbf y}
\newcommand{\zv}{\mathbf z}
\newcommand{\Cels}{$^{\circ}$C}

\newcommand{\oneD}{\text{1D}}
\newcommand{\twoD}{\text{2D}}

\usepackage{times}

\usepackage{color}

\usepackage[colorlinks,bookmarks=false,citecolor=blue,linkcolor=red,urlcolor=blue]{hyperref}
\usepackage{verbatim}

\def\be{\begin{equation}}
\def\ee{\end{equation}}
\def\bea{\begin{eqnarray}}
\def\eea{\end{eqnarray}}
\def\bs{\boldsymbol}
\def\vec{\mathbf}
\def\mc{\mathcal}

\begin{document}

\title{Hindered magnetic order from mixed dimensionalities in CuP$_{2}$O$_{6}$}

\author{R. Nath}
\email{rnath@iisertvm.ac.in}

\author{K. M. Ranjith}
\affiliation{School of Physics, Indian Institute of Science
Education and Research Thiruvananthapuram-695016, India}

\author{J. Sichelschmidt}
\author{M. Baenitz}
\affiliation{Max Planck Institute for Chemical Physics of Solids, 01187 Dresden, Germany}

\author{Y. Skourski}
\affiliation{Dresden High Magnetic Field Laboratory, Helmholtz-Zentrum Dresden-Rossendorf, 01314 Dresden, Germany}

\author{F.~Alet}
\affiliation{Laboratoire de Physique Th\'eorique, Universit\'e de Toulouse and CNRS, UPS (IRSAMC), 31062 Toulouse, France}

\author{I. Rousochatzakis}
\email{i.rousochatzakis@ifw-dresden.de}
\affiliation{Institute for Theoretical Solid State Physics, IFW Dresden, 01171 Dresden, Germany}

\author{A. A. Tsirlin}
\email{altsirlin@gmail.com}
\affiliation{Max Planck Institute for Chemical Physics of Solids, 01187 Dresden, Germany}
\affiliation{National Institute of Chemical Physics and Biophysics, 12618 Tallinn, Estonia}


\begin{abstract}
We present a combined experimental and theoretical study of the spin-$\frac12$ compound CuP$_2$O$_6$ that features a network of two-dimensional (2D) antiferromagnetic (AFM) square planes, interconnected via one-dimensional (1D) AFM spin chains. Magnetic susceptibility, high-field magnetization, and electron spin resonance (ESR) data, as well as microscopic density-functional band-structure calculations and subsequent quantum Monte-Carlo simulations, show that the coupling $J_{\sf 2D}\simeq 40$~K in the layers is an order of magnitude larger than $J_{\sf 1D}\simeq 3$~K in the chains. Below $T_N\simeq 8$~K, CuP$_2$O$_6$ develops long-range order, as evidenced by a weak net moment on the 2D planes induced by anisotropic magnetic interactions of Dzyaloshinsky-Moriya type. A striking feature of this 3D ordering transition is that the 1D moments grow significantly slower than the ones on the 2D units, which is evidenced by the persistent paramagnetic ESR signal below $T_N$. Compared to typical quasi-2D magnets, the ordering temperature of CuP$_2$O$_6$ $T_N/J_{\sf 2D}\simeq 0.2$ is unusually low, showing that weakly coupled spins sandwiched between 2D magnetic units effectively decouple these units and impede the long-range ordering.
\end{abstract}
\maketitle

\section{\textbf{Introduction}}
According to the central theorem by Mermin and Wagner,\cite{mermin1966} low-dimensional Heisenberg magnets with sufficiently short-ranged isotropic interactions do not order magnetically at any finite temperature in one or two spatial dimensions. The standard way in which real systems eventually evade this theorem is via weak inter-plane or inter-chain interactions which, no matter how weak, are enough to mitigate the entropic fluctuations and stabilize long range order (LRO) below a characteristic temperature $T_N$. By now, many aspects of this important ordering transition are theoretically well understood\cite{scalapino1975, schulz1996, irkhin1997, irkhin1998, irkhin1999, irkhin2000, bocquet2002, yasuda2005} and tested by many experiments.\cite{keimer1992,keren1993,kojima1997} For example, one of the key factors affecting the value of $T_N$ is the fluctuation strength of the low-dimensional units (planes or chains) and, hence, their dimensionality. This aspect is already exposed by mean-field theory,\cite{scalapino1975, schulz1996} which shows that for weakly coupled antiferromagnetic (AFM) planes (or chains), $T_N$ is given by
\be\label{eq:RPAa}
\chi^s(T_N(J'))=\frac{1}{z J'}~,
\ee
where $J'$ is the interlayer (resp. interchain) coupling, $\chi^s$ is the staggered susceptibility $\chi_{\sf 2D}^s$ (resp. $\chi_{\sf 1D}^s$), and $z=2$ (resp. $z=4$) is the coordination number.

Here, we study what happens when magnetic units of different dimensionality are brought together in a single material and, in particular, when successive AFM layers couple to each other only via intermediate AFM chains. This situation is realized in the spin-$\frac12$ compound CuP$_2$O$_6$, where a 2D sublattice with stronger magnetic couplings coexists with a 1D sublattice featuring weaker magnetic couplings.  Systems of this type have been scarcely studied in the past, both experimentally and theoretically, yet they raise a number of interesting and so far unresolved questions. What is the nature of the transition and the magnetic behavior of the two sublattices below $T_N$? Which energy scale determines the value of $T_N$?

The experiments presented below show clearly that the coexistence of two dimensionalities hinders the overall three-dimensional (3D) magnetic order in CuP$_2$O$_6$ and, additionally, leads to a very peculiar situation where the 1D sublattice remains partly paramagnetic below $T_N$. This physics can be accounted for by a mean-field treatment of the problem which, in particular, generalizes Eq.~(\ref{eq:RPAa}) to
\be\label{eq:RPAb}
\bar{\chi}^s(T_N(J'))=\frac{1}{z J'}~,
\ee
where $\bar{\chi}^s=\sqrt{\chi_{\sf 1D}^s\chi_{\sf 2D}^s}$ is the geometric mean of the two staggered susceptibilities. As we discuss below, the physical meaning of this equation can be understood in terms of an effective, temperature-dependent interlayer coupling $J'_{\eff,\perp}(T)$, mediated by the chains.

\begin{figure*}
\includegraphics{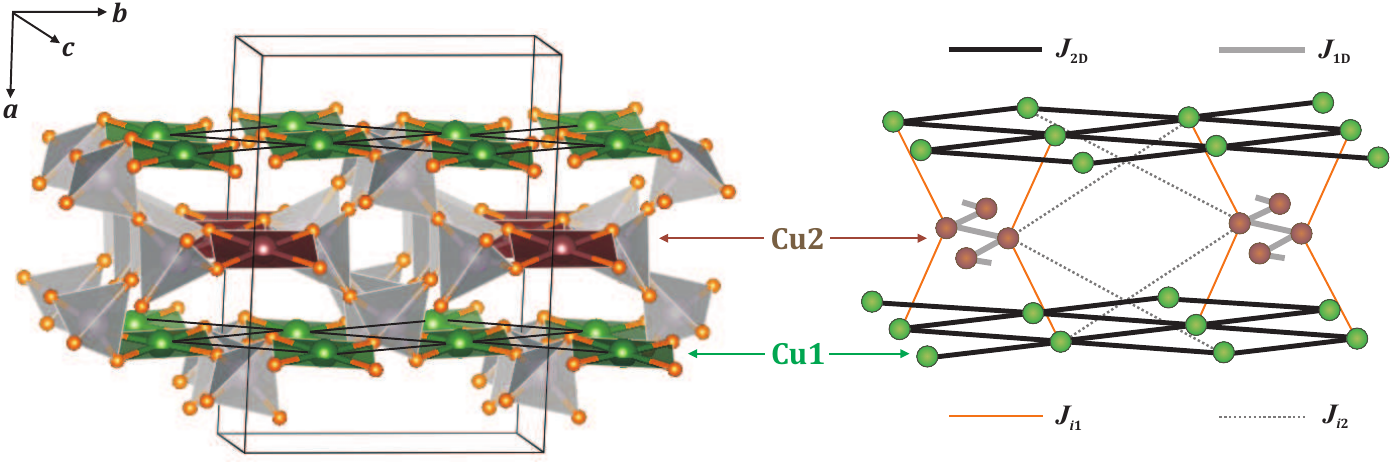}
\caption{\label{fig:str} Crystal structure of CuP$_2$O$_6$ (left) and the relevant magnetic model (right). Note two different Cu positions that form planar 2D units in the $bc$ plane (Cu1, green), and buckled chains stretched along the $c$ direction (Cu2, brown). For details of the Cu1 and Cu2 sublattices see Figs.~\ref{fig:sqlat} and~\ref{fig:chain}, respectively. Weak couplings $J_{i1}$ (thin solid line) and $J_{i2}$ (dotted line) are leading interactions between the sublattices. Crystal structures are visualized using the \texttt{VESTA} software.\cite{vesta}}
\end{figure*}

Regarding the crystal structure of CuP$_2$O$_6$, this compound features two crystallographically inequivalent positions of the magnetic Cu$^{2+}$ ions. Despite the similar local environment, which is well described by CuO$_4$ planar units, the two Cu sites possess different connectivities (Fig.~\ref{fig:str}). Shortest Cu1--Cu1 contacts form a planar sublattice, which is topologically equivalent to the square lattice: despite the lack of the overall four-fold symmetry, the monoclinic symmetry $C2/c$ renders all nearest-neighbor exchange pathways equivalent ($J_{\sf 2D}$). The Cu2 sites, on the other hand, form buckled chains stretched along the $c$ direction. The shortest Cu2--Cu2 distance along $b$ (8.09~\r A) is nearly twice larger than the intrachain separation of 4.86~\r A along $c$. Therefore, a 1D sublattice of Cu2 spins is expected (Fig.~\ref{fig:str}, right). In the following, we will verify this picture by a model treatment of the experimental data (Sec.~\ref{sec:mag}) and by a direct microscopic analysis (Sec.~\ref{sec:model}).

Some magnetic properties of CuP$_2$O$_6$ have been reported before,\cite{gunsser1989,belaiche2004} but no conclusive information regarding the nature of magnetism and magnetic order in this compound is available. While magnetic susceptibility and electron spin resonance (ESR) show signatures of a magnetic transition around 8~K, a neutron diffraction experiment failed to detect any magnetic reflections below this temperature.\cite{gunsser1989} In the following, we report a combined experimental and computational study of CuP$_2$O$_6$. In contrast to previous studies that focused only on phenomenological aspects of this material, we provide a detailed microscopic scenario of CuP$_2$O$_6$ and establish the combination of 1D and 2D magnetism taking place on different energy scales, thus leading to a magnetic ordering transition around 8~K with a remarkably low $T_N/J\simeq 0.2$. We argue that the coupling of 2D sublattices through the 1D sublattices is responsible for this effect. 

\section{\textbf{Methodology}}
Polycrystalline samples of CuP$_{2}$O$_{6}$ were prepared by the conventional solid-state reaction technique using CuO (Aldrich, $99.999$\%) and NH$_{4}$H$_{2}$PO$_{4}$ (Aldrich, $99.999$\%) as starting materials.\cite{bamberger1997}
The stoichiometric mixtures were ground and fired at 400~\Cels\ for 24~h. The product was not uniform in color. The sample was given subsequent firings at 450~\Cels\ and 500~\Cels\ for 24~h each. Phase purity of the samples were checked using powder x-ray diffraction (PANalytical) with CuK$_{\alpha}$ radiation ($\lambda$ = 1.54060~\r A). Powder x-ray diffraction was done after each firing. At 450~\Cels, a single-phase sample was obtained. The firing at 500~\Cels\ resulted in an impurity phase of Cu$_{2}$P$_{2}$O$_{7}$. Therefore, all our measurements were done on the 450~\Cels\ sample.

\begin{figure*}
\includegraphics{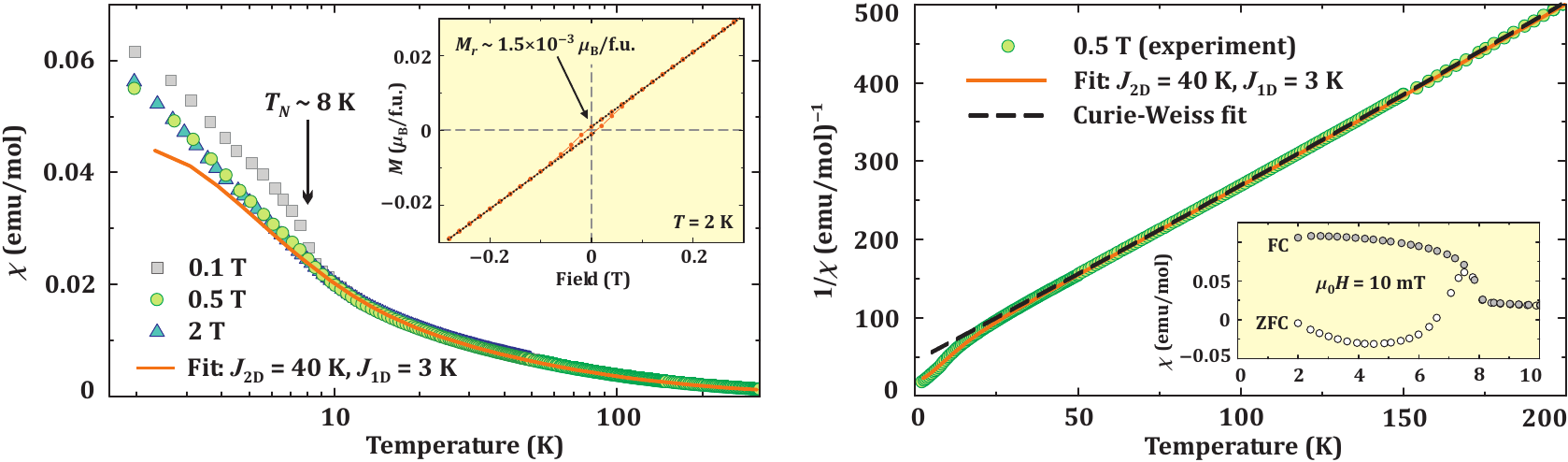}
\caption{\label{fig:chi} 
(Color online) Left panel: magnetic susceptibility ($\chi$) of CuP$_2$O$_6$ measured in the applied fields of 0.1~T, 0.5~T, and 2~T. The inset shows the magnetization curve at 2~K, with a tiny hysteresis and the remnant magnetization $M_r\simeq 1.5\times 10^{-3}$~$\mu_B$/f.u. Dotted lines are guides-for-the-eye. Right panel: inverse susceptibility ($1/\chi$). The inset shows the field-cooled and zero-field-cooled susceptibility measured at 10~mT. In both panels, the solid line is the QMC fit with the two-sublattice model, and the dashed line (right panel) is the Curie-Weiss fit, see text for details.
}
\end{figure*}
The magnetic susceptibility ($\chi$) was measured as a function of temperature (1.8~K $\leq T\leq$ 380~K) and at different applied magnetic fields ($H$). Zero-field-cooled (ZFC) and field-cooled (FC) magnetic susceptibility was measured at an applied field of 100~Oe and at low temperatures. Magnetization isotherms ($M$ vs. $H$) were also measured at $T=1.8-2.0$~K while both increasing and decreasing fields. All these magnetic measurements were performed using a SQUID magnetometer (Quantum Design MPMS) and the vibrating sample magnetometer option of a Quantum Design PPMS. Our continuous attempts to measure the heat capacity failed due to difficulties in making a hard sintered pellet. High-field data were collected using a pulsed magnet installed at the Dresden High Magnetic Field Laboratory. Details of the experimental procedure can be found elsewhere.\cite{tsirlin2009} 

The ESR measurements were carried out with a commercial ESR spectrometer at the X-band frequency (9.4~GHz) in the $4.2-300$~K temperature range.

The magnetic couplings in CuP$_2$O$_6$ were evaluated by density-functional (DFT) band-structure calculations performed in the \texttt{FPLO9.01-37}\cite{fplo} and \texttt{VASP5.2}\cite{vasp1,*vasp2} codes. The local density approximation (LDA)\cite{pw92} and generalized gradient approximation (GGA)\cite{pbe96} of the exchange-correlation potential were used. In addition to model analysis on top of the LDA band structure, we calculated isotropic exchange couplings and magnetic anisotropy parameters using GGA+$U$ calculations with the on-site Coulomb repulsion $U_d=8.5$~eV, on-site Hund's exchange $J_d=1$~eV, and the double-counting correction obtained in the atomic limit.\cite{janson2012,tsirlin2012b,berdonosov2013} The mean-field GGA+$U$ procedure provides a reasonable correction for effects of strong correlations in the Cu $3d$ shell, and facilitates a precise evaluation of the magnetic couplings. We estimate parameters of the following spin Hamiltonian:
\begin{equation}
 \mc{H}=\sum_{ij}\left( J_{ij}\vec{S}_i\!\cdot\!\vec{S}_j+\vec{D}_{ij}\!\cdot\!\vec{S}_i\!\times\!\vec{S}_j+\vec{S}_i\!\cdot\!\bs{\Gamma}_{ij}\!\cdot\!\vec{S}_j\right),
\label{eq:ham}\end{equation}
where the summations are over all interacting pairs of spins $\Sv_i$ and $\Sv_j$, $J_{ij}$ is the isotropic (Heisenberg) exchange coupling, $\Dv_{ij}$ is the antisymmetric Dzyaloshinsky-Moriya interaction, and $\bs{\Gamma_{ij}}$ is the symmetric and traceless portion of the anisotropy. Details of the computational procedure are described in Sec.~\ref{sec:model}, see also Refs.~\onlinecite{xiang2011,anisotropy}.

Thermodynamic properties were evaluated by quantum Monte-Carlo (QMC) simulations with the \texttt{loop}\cite{loop} and \verb|dirloop_sse|\cite{dirloop} algorithms implemented in the \texttt{ALPS} simulation package.\cite{alps} The simulations for CuP$_2$O$_6$ were performed on 3D $L\times L\times L/8$ finite lattices with periodic boundary conditions and $L$ up to 32. Additionally, properties of individual 1D and 2D units were calculated for low-dimensional 1D lattices with $L$ up to 8192 and 2D $L\times L$ lattices with $L$ up to 768. This choice of $L$ ensures that the uniform magnetic susceptibility, specific heat, and field-dependent magnetization reported in this work are free from finite-size effects for all relevant temperatures. In the case of the staggered susceptibility of the 2D lattice, we were able to reach size convergence down to $T/J_{\sf 2D}\simeq 0.21$, which is sufficient for fixing the value of $T_N$ from mean-field theory, see below.

\section{Results}
\subsection{Magnetization}
\label{sec:mag}
Bulk magnetic susceptibility $\chi$ as a function of temperature in different applied fields is shown in Fig.~\ref{fig:chi}. With decreasing temperature, $\chi(T)$ increases in a Curie-Weiss manner down to $20-25$~K and shows a faster increase at lower temperatures. The data measured in low fields additionally show a kink around $T_N\simeq 8$~K. Below this kink, the susceptibility increases abruptly, which indicates the formation of a net magnetic moment, as typical for systems with canted AFM order. The transition at 8~K is also visible in the low-field FC and ZFC data (right panel of Fig.~\ref{fig:chi}, inset) that eventually diverge around 7.5~K. At low temperatures, the ZFC susceptibility is negative, because the system orders in zero field, whereas the small external field applied after cooling the system through $T_N$ is insufficient for changing the direction of the net moment. The net magnetization of CuP$_2$O$_6$ is quite small. The $M(H)$ curve at 2~K (left panel of Fig.~\ref{fig:chi}, inset) shows a hardly visible hysteresis and yields the remnant magnetization of $M_r\simeq 1.5\times 10^{-3}$~$\mu_B$/f.u., as found from linear fits of the $M(H)$ data between 0.1 and 0.5~T.

We carefully checked for a possible extrinsic origin of this weak net moment. First, several samples were measured, and all showed the same transition temperature and same net moment. Second, possible magnetic impurities in CuP$_2$O$_6$ were considered. Those are restricted to CuO and several Cu$^{2+}$ phosphates, but none of them show magnetic transitions around 8~K. For example, Cu$_2$P$_2$O$_7$, which is a typical impurity phase in CuP$_2$O$_6$ samples, orders antiferromagnetically at 27~K and lacks any net magnetic moment.\cite{janson2011} Finally, our ESR data (Sec.~\ref{sec:esr}) evidence an intrinsic change in CuP$_2$O$_6$ around 8~K. Note that previous studies also reported the formation of the weak net moment in CuP$_2$O$_6$ below 8~K.\cite{gunsser1989} In the following, the microscopic origin of this weak ferromagnetism will be elucidated (Sec.~\ref{sec:anisotropy}).

To fit the bulk susceptibility data at high temperatures, we use the expression
\begin{equation}
\chi=\chi_{0}+\frac{C}{T+\theta_{\CW}},
\label{eq:cw}\end{equation}
\newline
where $\chi_0$ is the temperature-independent contribution and consists of diamagnetism of the core electron shells ($\chi_{\text{core}}$) and van Vleck paramagnetism ($\chi_{\text{VV}}$) of the open shells of the Cu$^{2+}$ ions present in the sample. The second term is the Curie-Weiss (CW) law with the Curie constant $C=N_A\mu_{\eff}^2/3k_B$ and Curie-Weiss temperature $\theta_{\rm CW}$. Our fit in the temperature range from 100~K to 380~K yields $\chi_0\simeq-9.5\times 10^{-5}$~emu/mol, $C \simeq 0.470$~emu~K/mol, and $\theta_{\rm CW}\simeq 20$~K. The positive value of $\theta_{\rm CW}$ suggests that the dominant interactions are AFM in nature. The $C$ value corresponds to an effective moment $\mu_{\eff}\!=\!g\mu_B\sqrt{S(S+1)}\simeq 1.91\mu_B$, yielding $g\!\simeq\! 2.21$. This value matches the powder-averaged $\bar g\simeq 2.22$ from ESR (Sec.~\ref{sec:esr}). The temperature-independent contribution $\chi_0$ is below the core diamagnetic susceptibility $\chi_{\text{core}}\simeq -8.5\times 10^{-5}$~emu/mol expected for individual ions.\cite{selwood1956} Therefore, the van Vleck contribution is likely very small (we expect $\chi_{\text{VV}}\simeq 1.5\times 10^{-5}$~emu/mol according to Refs.~\onlinecite{motoyama1996,nath2005}), while the remaining discrepancy between $\chi_{\text{core}}$ and $\chi_0$ is due to the weak diamagnetic signal of the sample holder.

\begin{figure}
\includegraphics{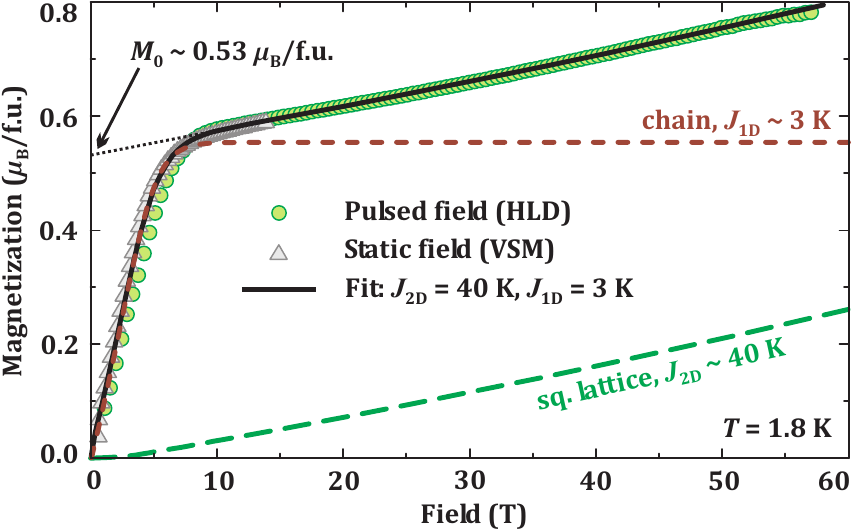}
\caption{\label{fig:mvsh} 
(Color online) Magnetization isotherm of CuP$_2$O$_6$ measured at $T\simeq 1.8$~K in static and pulsed fields. Solid line is the fit with the two-sublattice model, as described in the text, whereas short-dashed (brown) and long-dashed (green) lines denote individual contributions of the 1D and 2D sublattices, respectively. Dotted line shows the extrapolation of the linear region to zero field.}
\end{figure}

The magnetization of CuP$_2$O$_6$ increases with the field and bends around 6~T (Fig.~\ref{fig:mvsh}). In higher fields, the magnetization continues increasing, yet with a much lower slope, and shows linear behavior up to at least 55~T.\footnote{Here, we used the low-field data from static fields to scale the high-field data from pulsed fields. Slight discrepancies at low fields may be due to a minor difference in sample temperature and due to dynamic effects, which are inherent to pulsed-field experiments.} The extrapolation of this linear trend to zero field yields $M_0\simeq 0.53$~$\mu_B$/f.u., which is about half of the full magnetization $M_s=gS\mu_B\simeq 1.11$~$\mu_B$/f.u. Therefore, half of the Cu$^{2+}$ spins are saturated in low fields, whereas the remaining Cu$^{2+}$ spins are strongly coupled and can not be saturated even at 55~T. This phenomenological analysis is well in line with our above picture of two different Cu sublattices in CuP$_2$O$_6$ with different energy scales.

We successfully fitted the data by a simple sum of the uniform-chain and square-lattice contributions, both simulated by QMC. Note that the $\chi(T)$ and $M(H)$ curves are described by the same set of parameters: $J_{\sf 2D}\simeq 40$~K, $J_{\sf 1D}\simeq 3$~K, $g\simeq 2.22$, and $\chi_0=-9.5\times 10^{-5}$~emu/mol, where $\chi_0$ has the same meaning as in Eq.~\eqref{eq:cw} and conforms to the value from the Curie-Weiss fit. The fitted $g$-value matches the powder-averaged $\bar g=2.22$ from ESR (Fig.~\ref{fig:esr2}), whereas $J_{\sf 2D}$ and $J_{\sf 1D}$ are in excellent agreement with the DFT results presented in Sec.~\ref{sec:model}. The saturation field of CuP$_2$O$_6$ is expected at $\mu_0H_s=4J_{\sf 2D}\times k_B/(g\mu_B)\simeq 105$~T, which is also confirmed in our QMC simulations.

Remarkably, the susceptibility of CuP$_2$O$_6$ does not reveal the low-dimensional AFM behavior of this compound. The susceptibility follows the CW law down to $25-30$~K and keeps increasing down to $T_N$, in contrast to expectations for a low-dimensional (i.e., purely 1D or purely 2D) AFM spin system. Indeed, the Cu2 spins in spin chains feature only weak magnetic couplings and remain in the paramagnetic state even in the temperature range $T\leq J_{\sf 2D}$, where the Cu1 spins develop a 2D short-range order, which in a simple 2D magnet will manifest itself by a susceptibility maximum around $T\simeq J_{\sf 2D}\simeq 40$~K. This effect is not seen in the susceptibility, because the susceptibility of CuP$_2$O$_6$ is dominated by the paramagnetic contribution of the Cu2 spins. On the other hand, the short-range order in individual sublattices could be pinpointed by the specific heat data. Our heat-capacity measurements were so far unsuccessful. Nevertheless, we demonstrate the effect clearly in the simulated specific heat, see Fig.~\ref{fig:heat} and Sec.~\ref{sec:LRO}.

\begin{figure}
\includegraphics{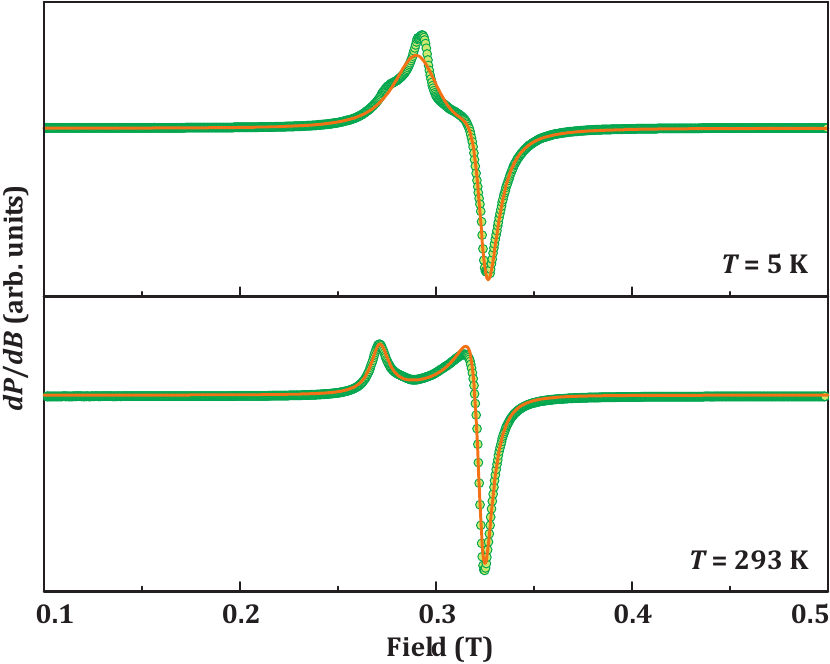}
\caption{\label{fig:esr1}
(Color online) ESR spectra measured at 5~K and 293~K and fits with powder-averaged Lorentzian lines assuming the easy-plane anisotropy. Note that at low temperatures the shape of the spectral line becomes more complex than at high temperatures, and additional terms beyond the single powder-averaged line may be required for a good fit.
}
\end{figure}

\subsection{Electron spin resonance}\label{sec:esr}
\begin{figure}
\includegraphics{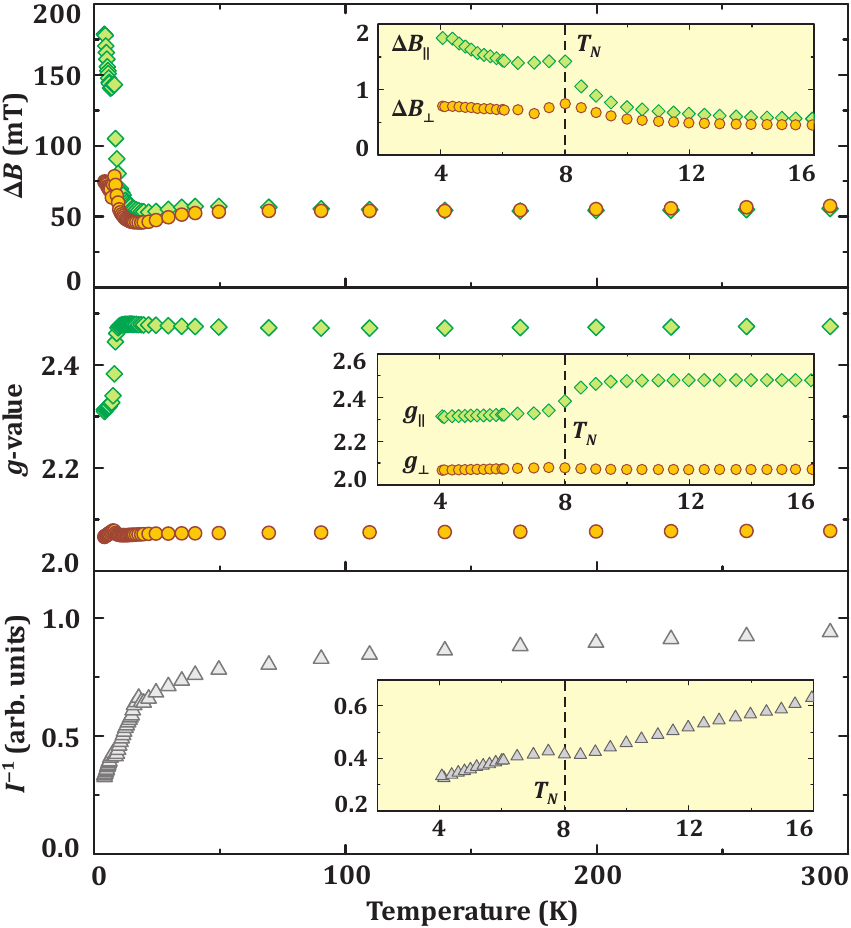}
\caption{\label{fig:esr2}
(Color online) Temperature dependence of the ESR spectral parameters: linewidth ($\Delta B$, top panel), $g$-values (middle panel), and inverse line intensity ($I^{-1}$, bottom panel). Circles and diamonds denote the components perpendicular and parallel to the field, respectively.}
\end{figure}

The ESR spectra of CuP$_2$O$_6$ can be fitted with a powder-averaged Lorentzian line for the case of easy-plane anisotropy of the $g$-tensor. Individual lines corresponding to the Cu1 and Cu2 positions could not be distinguished (Fig.~\ref{fig:esr1}). The easy-plane anisotropy implies two different $g$-values, $g_{\|}$ and $g_{\perp}$, respectively, which are evaluated from the line positions. We also monitor the linewidth ($\Delta B$) and the line intensity ($I$) as a function of temperature. Above 20~K, the $g$-values and $\Delta B$ are nearly temperature-independent, while $I$ slightly decreases following the decreasing magnetic susceptibility $\chi$ (Fig.~\ref{fig:esr2}). At lower temperatures, we observed a conspicuous effect at 8~K, where all parameters of the ESR lines change abruptly. Such abrupt changes are expected at a magnetic ordering transition, where the paramagnetic ESR signal usually disappears. However, in our case the ESR signal could be also observed below 8~K, yet with a different resonance field $g_{\|}$. 

The persistence of the paramagnetic ESR signal below $T_N$ is rooted in the different behavior of the two sublattices in CuP$_2$O$_6$. The standard situation in AFM resonance~\cite{Kittel1951,KefferKittel1952,MnF2, Zvyagin2010} is that below $T_N$ the resonance frequencies $\omega_{\sf res}$ shift far away from the paramagnetic value $\omega_0=g\mu_B H$, owing to the presence of anisotropy and the buildup of strong local exchange fields. This is exactly what we expect here for the Cu1 spins, since their moments grow quickly below $T_N$ and the anisotropy field is effectively induced by the DM interactions (or equivalently by the weak ferromagnetic moments, see Sec.~\ref{sec:anisotropy} below). The Cu2 spins on the chains, on the other hand, have a qualitatively different behavior: (i) there is no DM anisotropy along the chains due to inversion symmetry, and (ii) the moments grow significantly slower than those on Cu1 in the layers, because thermal fluctuations remain strong in a finite $T$ range $J_{\sf 1D} \lesssim T \le T_N$. Therefore, the persistence of the paramagnetic ESR signal below $T_N$ is essentially a fingerprint of this partly paramagnetic behavior of Cu1 spins. However, their resonance field, in particular $g_{\|}$, is modified because of internal fields developed in CuP$_2$O$_6$ below $T_N$.

\subsection{Microscopic magnetic model}\label{sec:model}
\subsubsection{Isotropic part}
\begin{figure}
\includegraphics{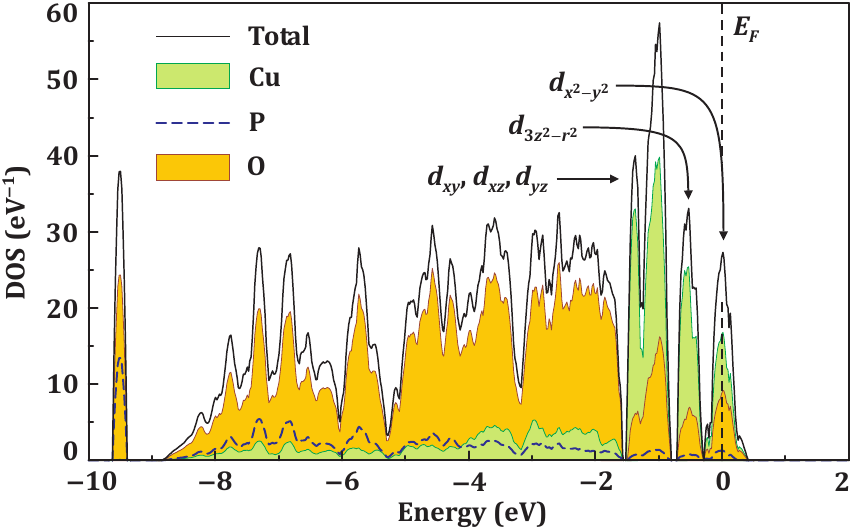}
\caption{\label{fig:dos}
(Color online) LDA density of states (DOS) for CuP$_2$O$_6$. The Fermi level ($E_F$) is at zero energy. Note well-defined crystal-field levels of Cu$^{2+}$ in the planar CuO$_4$ local environment.}
\end{figure}

Isotropic exchange couplings $J_i$ can be obtained from the dispersions of relevant LDA bands. The LDA band structure of CuP$_2$O$_6$ (Fig.~\ref{fig:dos}) resembles that of other Cu$^{2+}$ oxides and phosphates.\cite{janson2011,janson2012,tsirlin2010} The broad oxygen band between $-9$ and $-1.5$~eV is followed by mixed Cu~$3d$ -- O $2p$ bands above $-1.5$~eV. These Cu $3d$ states show additional structure related to the crystal-field levels of Cu$^{2+}$. In a local coordinate frame defined by the $x$ and $y$ axes directed along Cu--O bonds in the CuO$_4$ plane and the $z$ axis perpendicular to this plane, the bands between $-1.5$~eV and $-0.8$~eV have the $xy,yz$, and $xz$ origin. Narrow bands around $-0.5$~eV are due to the $3z^2-r^2$ orbitals, and the bands straddling the Fermi level have the $x^2-y^2$ origin. The latter band complex consists of four bands according to four Cu atoms in the primitive cell, two for each of the crystallographic positions. Note that the LDA band structure of CuP$_2$O$_6$ is metallic, because correlation effects, which are responsible for opening the band gap, are heavily underestimated in LDA. The missing part of the correlations is introduced in the model analysis or using the mean-field DFT+$U$ method. The DFT+$U$ calculations yield the large band gap of about 3.7~eV and magnetic moments of 0.9~$\mu_B$ on the Cu atoms.

\begin{figure}
\includegraphics{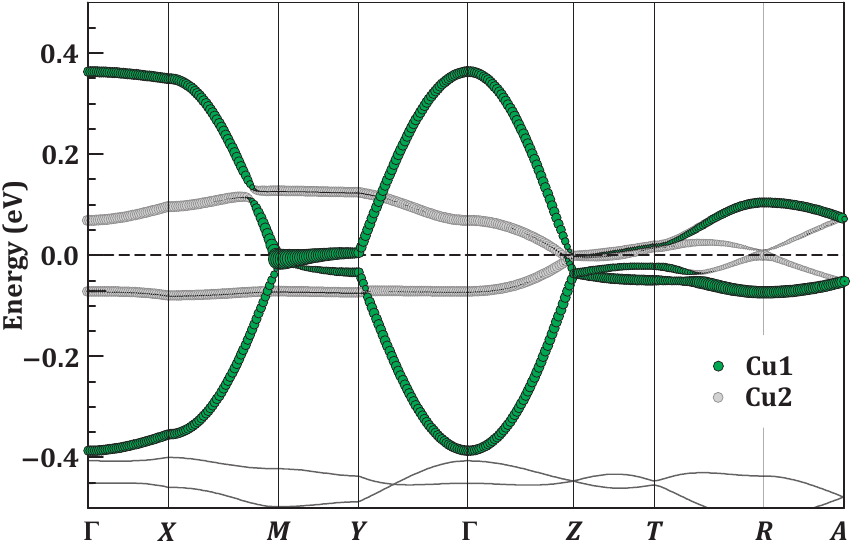}
\caption{\label{fig:bands}
(Color online) LDA band structure of CuP$_2$O$_6$ in the vicinity of the Fermi level. Green (dark) and gray (light) circles show the contributions of Cu1 and Cu2 $d_{x^2-y^2}$ orbitals, respectively. Note the different bandwidths that indicate different energy scales within the Cu1 and Cu2 sublattices. The $k$ path is defined as follows: $\Gamma(0,0,0)$, $X(0.5,0,0)$, $M(0.5,0.5,0)$, $Y(0,0.5,0)$, $Z(0,0,0.5)$, $T(0.5,0,0.5)$, $R(0.5,0.5,0.5)$, $A(0,0.5,0.5)$, where the coordinates are given in units of the respective reciprocal lattice parameters.}
\end{figure}

The different energy scales of the Cu1 and Cu2 sublattices are clearly visible from the different widths of the energy bands related to the Cu1 and Cu2 sites (Fig.~\ref{fig:bands}). The dispersions of the LDA bands are quantified using Wannier functions based on the Cu $d_{x^2-y^2}$ orbital character.\cite{wannier} The resulting hopping parameters $t_i$ are introduced into a Hubbard model with the effective on-site Coulomb repulsion $U_{\eff}\simeq 4.5$~eV.\cite{janson2011,schmitt2010,tsirlin2010} As \mbox{$t_i\ll U_{\eff}$}, the limit of strong correlations is well-justified, and lowest-lying (magnetic) excitations can be described with the Heisenberg model parametrized by AFM exchanges $J_i^{\AFM}=4t_i^2/U_{\eff}$. The resulting values are listed in Table~\ref{tab:hoppings} and confirm our qualitative assessment of the CuP$_2$O$_6$ structure in terms of two different magnetic sublattices. The largest AFM coupling $J_{\sf 2D}^{\AFM}\simeq 80$~K is found in the Cu1 sublattice (2D, square lattice). The couplings $J_{\sf 1D}^{\AFM}\simeq 23$~K between the Cu2 sites are much weaker and form the chain sublattice (1D). The weak couplings $J_{i1}$ and $J_{i2}$ link the sublattices. 

\begin{table}
\caption{\label{tab:hoppings}
Isotropic magnetic couplings in CuP$_2$O$_6$: Cu--Cu distances $d_{\text{Cu--Cu}}$ (in~\r A), LDA hoppings $t_i$ (in~meV), and ensuing AFM exchanges $J_i^{\AFM}=4t_i^2/U_{\eff}$ (in~K), where we use $U_{\eff}=4.5$~eV for the effective on-site Coulomb repulsion. The total exchange couplings $J_i$ (in~K) are obtained from GGA+$U$ calculations with $U_d=8.5$~eV and $J_d=1$~eV.}
\begin{ruledtabular}
\begin{tabular}{cccrr}
           & $d_{\text{Cu--Cu}}$ & $t_i$    & $J_i^{\AFM}$ & $J_i$ \\
 $J_{\sf 2D}$     &      6.264          & $-0.088$ &     80       &  38   \\
 $J_{\sf 1D}$     &      4.864          & $-0.047$ &     23       &   4   \\
 $J_{i1}$  &      3.303          & $-0.015$ &     2.3      &  0.5  \\
 $J_{i2}$  &      8.454          & $-0.017$ &     3.0      &  0.2  \\
\end{tabular}
\end{ruledtabular}
\end{table}

The LDA results are supported by DFT+$U$ calculations. Here, we obtain individual couplings $J_i$ by calculating total energies of several collinear spin configurations and mapping these energies onto a Heisenberg model. This way, we find $J_{\sf 2D}\simeq 38$~K and $J_{\sf 1D}\simeq 4$~K in excellent agreement with the experimental estimates of $J_{\sf 2D}\simeq 40$~K and $J_{\sf 1D}\simeq 3$~K. Note that all $J_i$ values are reduced compared to $J_i^{\AFM}$. This reduction is typical for cuprates, where even long-range couplings feature non-negligible ferromagnetic (FM) contributions to the exchange.\cite{janson2009,lebernegg2011} However, the total couplings $J_i$ are rather different from those found in Cu$^{2+}$ compounds with similar structural units. 

\begin{figure}
\includegraphics{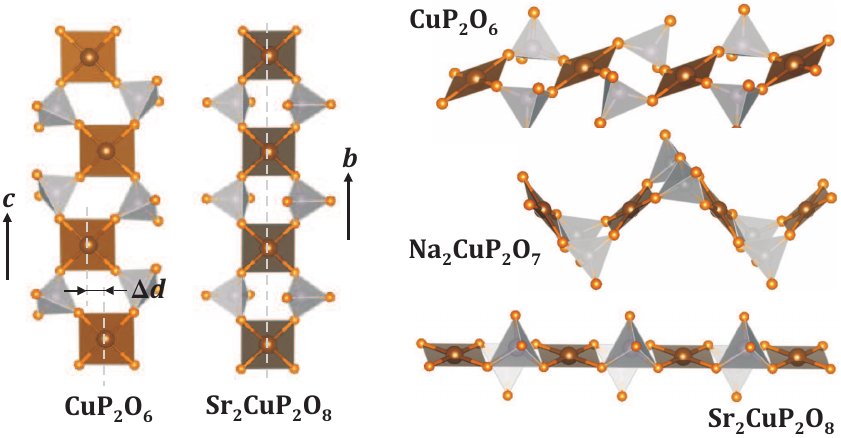}
\caption{\label{fig:chain}
(Color online) Comparison of the chain sublattices in CuP$_2$O$_6$, Na$_2$CuP$_2$O$_7$, and Sr$_2$CuP$_2$O$_8$. Note the flat and linear chain in Sr$_2$CuP$_2$O$_8$ (large exchange $J\simeq 140$~K), buckled linear chain in Na$_2$CuP$_2$O$_7$ (moderate exchange $J\simeq 27$~K), and buckled zigzag chain in CuP$_2$O$_6$ (weak exchange $J_{\sf 1D}\simeq 4$~K).}
\end{figure}

The geometry of the Cu2 spin chain resembles double-bridged spin chains in other Cu$^{2+}$ phosphates (Fig.~\ref{fig:chain}). In Sr$_2$CuP$_2$O$_8$ and K$_2$CuP$_2$O$_7$, the coupling through the double bridges of PO$_4$ tetrahedra exceeds 100~K,\cite{belik2004,nath2005,johannes2006,nath2008b} whereas in CuP$_2$O$_6$ it is only 4~K, well below $J_{\sf 2D}$, even though the Cu--Cu distance for $J_{\sf 2D}$ is much longer (Table~\ref{tab:hoppings}). The very low value of $J_{\sf 1D}$ is related to two geometrical effects. First, the chain is buckled, so that the CuO$_4$ plaquettes of the neighboring Cu2 atoms do not lie in the same plane. Second, these plaquettes are displaced with respect to each other, thus leading to a displacement of $\Delta d=0.88$~\r A compared to $\Delta d=0$ in Sr$_2$CuP$_2$O$_8$ and K$_2$CuP$_2$O$_7$. Both effects reduce the orbital overlap along the Cu--O--O--Cu pathways and, therefore, impede the magnetic coupling. The role of these effects is additionally illustrated by Na$_2$CuP$_2$O$_7$,\cite{lebernegg2011} where the chains are strongly buckled,\footnote{Note that the way of buckling is different from CuP$_2$O$_6$} but all Cu atoms lie on the same line as in Sr$_2$CuP$_2$O$_8$. The resulting coupling of 27~K is intermediate between $100-150$~K in linear and flat chains (Sr$_2$CuP$_2$O$_8$, K$_2$CuP$_2$O$_7$) and 4~K in buckled zigzag chains (CuP$_2$O$_6$).

Regarding the Cu1 sublattice (Fig.~\ref{fig:sqlat}), its structural unit resembles Ba$_2$CuGe$_2$O$_7$, where the leading in-plane coupling is around 11~K,\cite{zheludev1996} about 4 times lower than $J_{\sf 2D}$ in CuP$_2$O$_6$. This difference can be traced back to different orientations of the neighboring CuO$_4$ plaquettes. The structure of Ba$_2$CuGe$_2$O$_7$ features notable in-plane rotations of the CuO$_4$ plaquettes with the rotation angle of $\varphi\simeq 40^{\circ}$. In CuP$_2$O$_6$, the plaquettes are connected via single tetrahedra instead of the double Ge$_2$O$_7$ groups in Ba$_2$CuGe$_2$O$_7$. Therefore, the rotations are eliminated ($\varphi\simeq 0$), individual Cu--O bonds are nearly co-aligned, and a more efficient Cu--O--O--Cu superexchange occurs.

The Cu1 and Cu2 sublattices are stacked on top of each other and alternate along the $a$ direction. No substantial Cu2--Cu2 interchain coupling in the $bc$ plane has been found. The three-dimensional (3D) magnetic order should be rather driven by the Cu1--Cu2 couplings $J_{i1}$ and $J_{i2}$ connecting the sublattices. These weak couplings are frustrated, because both are weakly AFM and together with $J_{\sf 2D}$ form triangles (Fig.~\ref{fig:str}, right). However, the magnitude of the frustration is somewhat difficult to estimate, as both $J_{i1}$ and $J_{i2}$ are below 1~K and, thus, well within the typical error bar of DFT estimates. A perfect frustration is very unlikely, because $J_{i1}$ and $J_{i2}$ are not related by symmetry. While a competition between these couplings may be detrimental for the 3D magnetic order, magnetic ordering is far from being trivial even in a non-frustrated system of this type (see Sec.~\ref{sec:LRO} for further discussion).

\subsubsection{Magnetic anisotropy}\label{sec:anisotropy}
Let us now consider anisotropic terms in the spin Hamiltonian [Eq.~\eqref{eq:ham}]. The values of $\Dv_i$ and $\bs{\Gamma}_i$ are obtained from DFT+$U$ calculations, similar to the isotropic couplings $J_i$. The main difference between the evaluation of $J_i$ and $\Dv_i$ or $\bs{\Gamma_i}$ is the spin-orbit coupling that relates the spin direction to a certain crystallographic direction. For the DM vectors $\Dv_i$ and non-diagonal terms of the $\bs{\Gamma_i}$ tensor, orthogonal rather than collinear spin configurations should be used. We evaluate the anisotropy terms for the leading exchange pathway ($J_{\sf 2D}$) only. Other couplings are weak, hence their anisotropies are likely well below 1~K and, thus, can not be addressed by our present computational method. Note also that the neighboring Cu2 sites are linked by an inversion symmetry that eliminates the DM anisotropy for $J_{\sf 1D}$.

\begin{figure}
\includegraphics{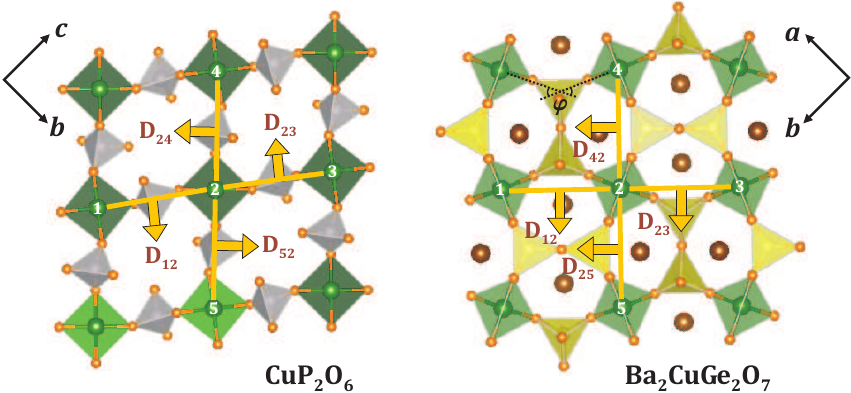}
\caption{\label{fig:sqlat}
(Color online) Comparison of the 2D sublattices in CuP$_2$O$_6$ (left) and Ba$_2$CuGe$_2$O$_7$ (right). Large arrows show the spatial arrangement of the DM vectors (only in-plane components), which are uniform in Ba$_2$CuGe$_2$O$_7$ and alternate in CuP$_2$O$_6$. Dotted lines denote the rotation of the neighboring CuO$_4$ plaquettes by an angle $\varphi\simeq 40^{\circ}$. In CuP$_2$O$_6$, this rotation is very weak ($\varphi\simeq 0^{\circ}$), hence the Cu--O bonds are nearly co-aligned, and the isotropic coupling $J_{\sf 2D}\simeq 40$~K is much larger than $J\simeq 11$~K in Ba$_2$CuGe$_2$O$_7$. In CuP$_2$O$_6$, the crystallographic axes $b$ and $c$ match, respectively, the directions $y$ and $z$ for the DM vectors.
}
\end{figure}

The in-plane component of the DM vector ($\Dv$) is nearly perpendicular to the Cu--Cu bonds and amounts to $D^{yz}\simeq 0.7$~K.\cite{foot1} The out-of-plane component is not forbidden by symmetry but it is much smaller, $D^x\simeq 0.1$~K. The symmetric anisotropy $\bs{\Gamma}$ is also very weak. All components of the $\bs{\Gamma}$ tensor are below 0.1~K. The spatial arrangement of the DM vectors can be inferred from the crystallographic symmetry of the Cu1 sublattice. The Cu1 atom is located in the inversion center. Therefore, the directions of the DM vectors alternate on the neighboring bonds as follows (see Fig.~\ref{fig:sqlat}, left): $\Dv_{12}=-\Dv_{23}$ and $\Dv_{52}=-\Dv_{24}$. The bonds 1--2 and 2--4 are related by the $c$ glide plane (reflection followed by the $c/2$ translation), which is perpendicular to the $b$ axis and results in $D_{12}^y=D_{24}^y$, $D_{12}^z=-D_{24}^z$. As $|D_{12}^y|\simeq |D_{12}^z|$, the DM vectors change their directions for about 90$^{\circ}$ between the adjacent Cu--Cu bonds and remain roughly perpendicular to the Cu--Cu bonds on the square lattice. 

These results should be compared to Ba$_2$CuGe$_2$O$_7$, where the in-plane component of the DM vector is also perpendicular to the Cu--Cu bonds ($D^{xy}\simeq 1.9$~K),\cite{zheludev1996} and additionally a weak out-of-plane component $D^z\simeq 0.08$~K is present\cite{muhlbauer2011,*muhlbauer2012,chovan2013} (note the exchanged crystallographic directions, because the $bc$ plane of CuP$_2$O$_6$ corresponds to the $ab$ plane of Ba$_2$CuGe$_2$O$_7$). Compared to CuP$_2$O$_6$ with $|\Dv|/J\simeq 0.018$, Ba$_2$CuGe$_2$O$_7$ features an order-of-magnitude larger DM anisotropy $|\Dv|/J\simeq 0.18$. This difference is again related to the in-plane rotations of the CuO$_4$ plaquettes (Fig.~\ref{fig:sqlat}) that reduce electron hoppings between orbitals of the same symmetry ($d_{x^2-y^2}\rightarrow d_{x^2-y^2}$ responsible for $J$) and amplify the hoppings between orbitals of different symmetry ($d_{x^2-y^2}\rightarrow d_{yz},d_{xz},d_{xy}$ responsible for different components of $\Dv$).\cite{mazurenko2008} 

Another difference between CuP$_2$O$_6$ and Ba$_2$CuGe$_2$O$_7$ pertains to the mutual arrangement of the DM vectors. The lack of inversion symmetry in Ba$_2$CuGe$_2$O$_7$ implies that the DM vectors on neighboring bonds do not alternate, hence $D_{12}^{xy}=D_{23}^{xy}$ and $D_{25}^{xy}=D_{42}^{xy}$.\cite{chovan2002} This uniform arrangement of the DM vectors leads to a weak spin canting and ensuing helical modulation seen experimentally.\cite{zheludev1996} In CuP$_2$O$_6$, the canting will rather produce a net magnetic moment, similar to the low-temperature phase of La$_2$CuO$_4$, where the DM vectors show the same staggered arrangement and lead to a canted AFM order.\cite{coffey1990,*coffey1991,shekhtman1992,*shekhtman1993,koshibae1994} 

Using the theoretical framework\cite{koshibae1994} established for La$_2$CuO$_4$, we conclude that the Cu1 sublattice of CuP$_2$O$_6$ forms a canted AFM order in the $bc$ plane, with the net moment pointing approximately along the $a$ direction\cite{foot1} perpendicular to the DM vectors. The primary spin direction lies in the $bc$ plane and depends on a subtle balance between the in-plane components of $\bs{\Gamma}$, which are difficult to determine from DFT. The canting angle is\cite{koshibae1994} $\theta=\frac12\tan^{-1}(|\Dv|/\sqrt 2J_{\sf 2D})\simeq 0.35^{\circ}$, given that the symmetric anisotropy $\bs{\Gamma}$ is neglected ($|\Gamma_{\alpha\beta}|\leq 0.1$~K). The resulting net moment $M_r=M_0\sin\theta\simeq 3.7\times 10^{-3}$~$\mu_B$/Cu1$\simeq 1.8\times 10^{-3}$~$\mu_B$/f.u., where $M_0\simeq 0.6$~$\mu_B$ is the ordered magnetic moment in a quasi-2D system of weakly coupled square planes,\cite{sandvik1997} and we took into account that the Cu2 spins do not contribute to the net moment. This estimate is in excellent agreement with the experimental value of $M_r\simeq 1.5\times 10^{-3}$~$\mu_B$/f.u. (Sec.~\ref{sec:mag}). This way, we confirm microscopically that the symmetry of the Cu1 sublattice of CuP$_2$O$_6$ entails the canted AFM order. Therefore, the experimental observation of the net moment below 8~K is a direct evidence for the long-range order of spins in the Cu1 sublattice.

We also attempted to assess qualitatively the anisotropy of the $g$-tensor. The interaction with the magnetic field activates a further anisotropy contained in the Zeeman term of the Hamiltonian:
\begin{equation}
\mc{H}_{\text{mag}}=\mc{H}+\sum_i\mu_B\vec{S}_i\!\cdot\!\bs{g}_i\!\cdot\!\vec{H},
\end{equation}
where $\mc{H}$ is the spin Hamiltonian in zero field [Eq.~\eqref{eq:ham}], $\gv_i$ is the $g$-tensor at site $i$, and $\Hv$ is the applied magnetic field. ESR data indicate a sizable easy-plane anisotropy of the $g$-tensors of both Cu1 and Cu2. While a full DFT-based evaluation of the $g$-tensor is presently difficult, at least for periodic systems, we can estimate a closely related quantity by calculating orbital moments of Cu$^{2+}$ for different directions of the spin moment. This way, we obtain tensors $\vec{L}$ that are composed of orbital moments (in units of $\mu_B$) calculated for spins directed along $x,y$, and $z$:
\begin{gather*}
\Lv_{\text{Cu1}}=\left(
\begin{array}{ccc}
  0.214  & 0.005    & $-0.024$ \\
  0.001  & 0.101    & $-0.005$ \\
$-0.027$ & $-0.005$ & 0.092    \\
\end{array}
\right), \\
\Lv_{\text{Cu2}}=\left(
\begin{array}{ccc}
  0.211  & 0.001    & $-0.033$ \\
  0.001  & 0.097    & 0.003    \\
$-0.036$ & 0.001    & 0.106    \\
\end{array}
\right).
\end{gather*}
Larger orbital moments imply larger components of the $g$-tensor, because $g$ defines the transformation from the spin moment to the full moment, i.e., to the sum of spin and orbital moments. 

Our computed orbital moments are in agreement with the easy-plane anisotropy, as $L^{xx}>L^{yy}\simeq L^{zz}$. The hard axis of the Cu$^{2+}$ ion is directed (approximately) along the crystallographic $a$ direction perpendicular to the CuO$_4$ plane. This is similar to other Cu$^{2+}$ square-lattice compounds: for example, in Ba$_2$CuGe$_2$O$_7$ $g_{ab}\simeq 2.04$ and $g_c\simeq 2.47$.\cite{zheludev1997} The Cu1 and Cu2 sites feature very similar matrices $\Lv$. Therefore, their $g$-tensors are likely similar, in agreement with the experimental ESR data, where the signals of the Cu1 and Cu2 atoms could not be distinguished (Sec.~\ref{sec:esr}).

\subsection{Long-range magnetic order}\label{sec:LRO}
\subsubsection{QMC simulations}
Having established the relevant microscopic couplings in the Cu1 and Cu2 sublattices, we now address the problem of the long-range magnetic ordering in CuP$_2$O$_6$. This ordering should, at least to some extent, involve both sublattices, because the interactions between the Cu1 sublattices take place via the Cu2 atoms, and the other way around (see Fig.~\ref{fig:str}). Now, as discussed above, including both $J_{i1}$ and $J_{i2}$ leads to frustration. Then the problem is not amenable to a low-temperature QMC study owing to the notorious sign problem. On the other hand, frustration is not expected to affect the essential aspects of the 3D ordering since $J_{i1}\neq J_{i2}$. Therefore, in the following we shall neglect the smallest coupling $J_{i2}$ and consider the effect of $J_{i1}$ only. 

\begin{figure}
\includegraphics{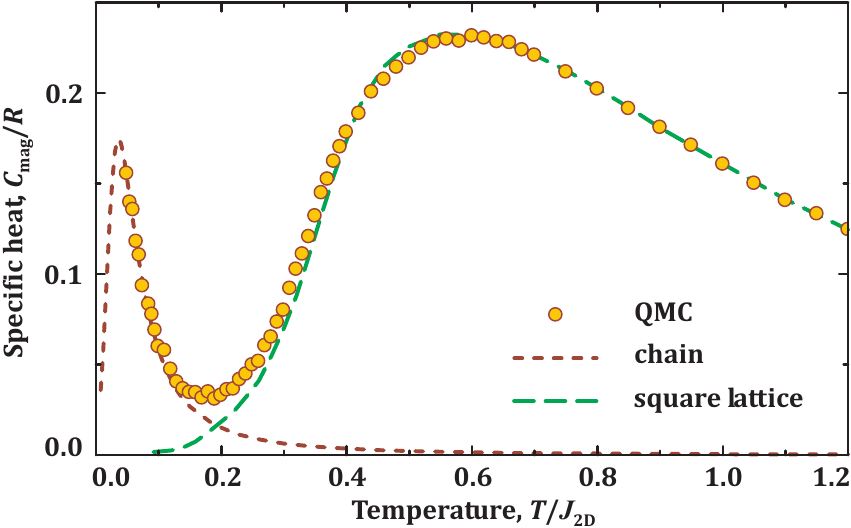}
\caption{\label{fig:heat}
(Color online) Simulated magnetic specific heat ($C_{\text{mag}}/R$) per lattice site for the spin lattice of CuP$_2$O$_6$ (circles) and individual contributions of the 1D and 2D sublattices shown by short-dashed and long-dashed lines, respectively. 
}
\end{figure}

We performed QMC simulations for a non-frustrated 3D spin lattice with $J_{\sf 1D}/J_{\sf 2D}=0.075$ and $J_{i1}/J_{\sf 2D}=0.01$, according to the DFT results (Table~\ref{tab:hoppings}) and the fits to the thermodynamic properties. Simulated magnetic specific heat shows two maxima arising from two spin sublattices of CuP$_2$O$_6$ (Fig.~\ref{fig:heat}). These maxima are well described by the contributions of individual chains and square planes and, thus, clearly indicative of the low-dimensional magnetic behavior. Surprisingly, no kinks or anomalies are seen down to at least $T/J_{\sf 2D}=0.1$, even though the experimental magnetic ordering temperature is $T_N/J_{\sf 2D}\simeq 0.2$. In fact, at low temperatures specific heat anomalies become diminutively small, because a very low amount of entropy is released at the magnetic transition. For example, Sengupta~\textit{et al.}\cite{sengupta2003} have shown that in quasi-2D systems the transition anomaly vanishes already at $T_N/J_{\sf 2D}\simeq 0.35$, which is also in agreement with the experimental studies.\cite{lancaster2007} Therefore, it is quite natural that we are unable to track the magnetic transition in the simulated specific heat.

QMC simulations are done for finite lattices that, as such, do not show LRO at any finite temperature. On the other hand, we did observe characteristic increase in the spin stiffness and staggered magnetization below $T/J_{\sf 2D}=0.3-0.4$. These observations are compatible with the fact that CuP$_2$O$_6$ undergoes LRO at $T_N/J_{\sf 2D}\simeq 0.2$. A detailed QMC study of the ordered state in this material lies beyond the scope of the present work and should be addressed in future studies. Presently, we restrict ourselves to a simple mean-field picture that clarifies the mechanism of the LRO in this type of spin system, where 1D and 2D spin sublattices coexist.

\begin{figure}
\includegraphics{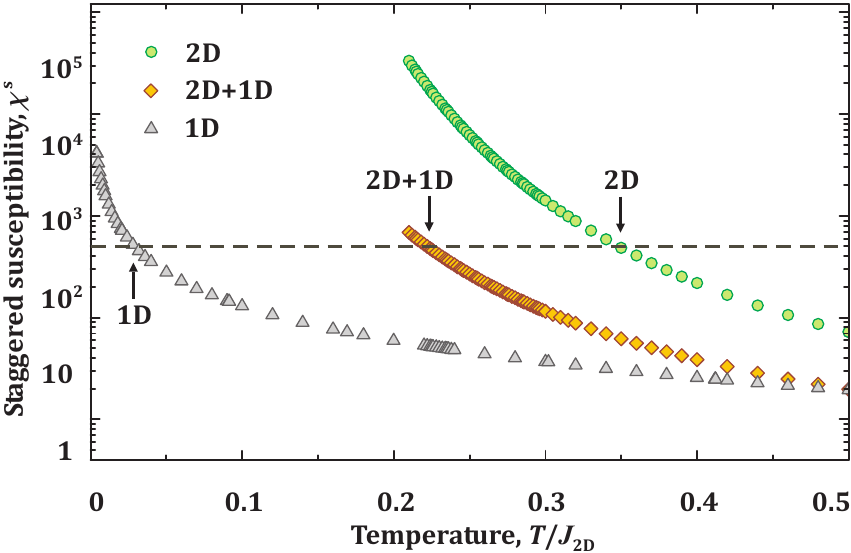}
\caption{\label{fig:mean-field}
(Color online) Staggered susceptibilities of the uniform spin chain (1D), square lattice (2D), and their combination, as observed in CuP$_2$O$_6$, where $\bar\chi^s=\sqrt{\chi^s_{\oneD}\chi^s_{\twoD}}$. Dashed line shows $(zJ_{i1}/J_{\sf 2D})^{-1}=50$ according to $J_{i1}/J_{\sf 2D}=0.01$ and $z=2$ in CuP$_2$O$_6$. The susceptibility of the uniform chain is multiplied by a factor of two to account for $z=4$ in a quasi-1D system. The truncation of the data at $T/J_{\sf 2D}=0.21$ marks the lowest temperature where the convergence with respect to the lattice size could be obtained for the 2D system.}
\end{figure}

\subsubsection{Mean-field theory}
\label{sec:mf}
\begin{figure}[!t]
\includegraphics[width=0.25\textwidth]{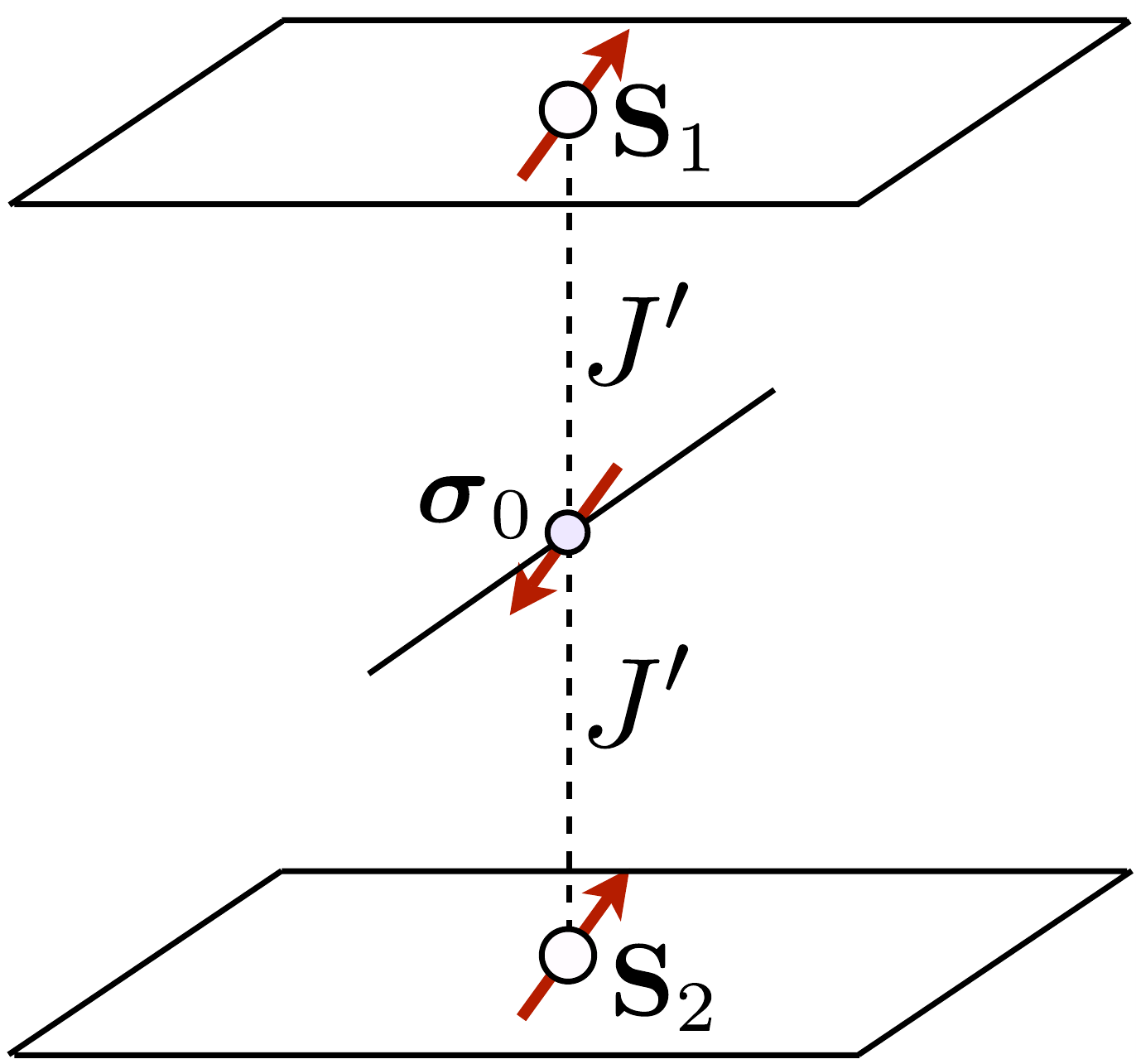}
\caption{\label{fig:MFT}
(Color online) A Cu2 spin chain sandwiched between two successive Cu1 square-plane antiferromagnets.}
\end{figure}

Here, we discuss the mean-field treatment of the coupling between the Cu1 and Cu2 spins. Figure~\ref{fig:MFT} shows two successive Cu1 square planes and a single Cu2 spin $\bs{\sigma}_0$ in between.  The exchange coupling between $\bs{\sigma}_0$ and the Cu1 spins $\vec{S}_1$ and $\vec{S}_2$ is denoted here by $J'$ (i.e. $J'\!=\!J_{i1}$), and is assumed to be much weaker than the intralayer coupling $J_{\sf 2D}$. Note also that we should keep the $J_{\sf 1D}$ coupling along the Cu2 chains, since $J_{\sf 1D}\simeq 4$~K in CuP$_2$O$_6$. We also assume an external field $\vec{H}_0$ which is conjugate to the order parameter, i.e., staggered along the Cu1 layers, along the Cu2 chains, and also along the $a$-direction, which is perpendicular to both the planes and the chains (see Fig.~\ref{fig:MFT}).

The physics that we want to capture by mean-field theory refers to the wide high-temperature range $T\!\gg\!J'$. In this range, the fluctuations of $\bs{\sigma}_0$ are essentially identical with those of a paramagnetic spin-$\frac12$ in the field $\vec{H}_0$. These fluctuations induce an effective, $T$-dependent coupling between the Cu1 layers, $J_{{\sf eff},\perp}(T)$, which (no matter how small) induces the 3D ordering transition of the Cu1 spins at $T_N$. In principle, the Cu2 spins should also become polarized at $T_N$, but the polarization is very weak as long as the internal fields $(\propto\!J')$ exerted from the Cu1 spins are much smaller than $T$. This means that Cu2 spins partly retain a paramagnetic behavior in a finite $T$-range below $T_N$. 

In the mean-field treatment of $J'$, the effective fields on the spins $\vec{S}_1$, $\vec{S}_2$ and $\bs{\sigma}_0$ read, in units of $k_B\!=\!g\mu_B\!=\!1$: $\vec{H}_{\eff,1}\!=\!\vec{H}_{\eff,2}\!=\!\vec{H}_0\!-\!z J'\langle \bs{\sigma}_0 \rangle$ and $\vec{H}_{\eff,0}\!=-\!\vec{H}_0\!-\!z J'\langle \vec{S}_{1} \rangle$, where $\langle\vec{S}_1\rangle\!=\!\langle\vec{S}_2\rangle$ and $z\!=\!2$, since each Cu1 spin interacts with two Cu2 spins (one above and one below the plane), and vice versa. In the linear response regime, we have
\be
\langle\sigma_0\rangle \!=\! \chi_{\sf 1D}^s\left[ H_0\!-\!z J' \langle S_1\rangle \right],~
\langle S_1\rangle \!=\! \chi_{\sf 2D}^s \left[-H_0\!-\!z J' \langle\sigma_0\rangle \right],
\ee
where $\chi_{\sf 1D}^s$ and $\chi_{\sf 2D}^s$ are the exact staggered susceptibilities of the AFM spin-$\frac12$ chain and square lattice, respectively. Solving for $\chi_0^s\!\equiv\!-\langle\sigma_0\rangle/H_0$ and $\chi^s\!\equiv\!\langle S_1\rangle/H_0$, yields:
\be
\chi^s \!=\! \frac{1\!+\!z J'\chi_{\sf 1D}^s}{1\!-\!(z J')^2\chi_{\sf 1D}^s \chi_{\sf 2D}^s} \chi_{\sf 2D}^s,~~
\chi_0^s \!=\! \frac{1\!+\!z J'\chi_{\sf 2D}^s}{1-(zJ')^2\chi_{\sf 1D}^s\chi_{\sf 2D}^s}\chi_{\sf 1D}^s~.
\ee
At $T_N$, the susceptibilities diverge, leading to:
\be\label{eq:RPA1}
\sqrt{\chi_{\sf 1D}^s\chi_{\sf 2D}^s} \equiv \bar{\chi}^s = \frac{1}{z J'}~.
\ee
The last equality is the well-known RPA condition,\cite{scalapino1975,schulz1996,irkhin2000,yasuda2005} where $\bar{\chi}^s$ plays the role of a ``geometric mean susceptibility''. To probe its physical meaning we rewrite it as 
\be\label{eq:RPA2}
\chi_{\sf 2D}^s(T_N) \!=\! \frac{1}{z J_{{\sf eff},\perp}(T_N)},~~
J_{{\eff},\perp}(T)\!=\! z J'^2 \chi_{\sf 1D}^s(T),
\ee 
which is now the RPA condition for the case of AFM planes that are coupled directly by $J_{{\sf eff},\perp}(T)$. The latter can be thought of as an effective coupling, mediated by the (paramagnetic) fluctuations of the Cu2 spins and, as such, it is $T$-dependent (via $\chi_{\sf 1D}^s(T)$) and scales quadratically with $J'$.

Figure~\ref{fig:mean-field} shows $\bar\chi^s(T)$ for CuP$_2$O$_6$ and for two reference systems representing standard quasi-1D and quasi-2D regimes of weakly coupled spin chains and square planes, respectively. In the case of CuP$_2$O$_6$, we used $J_{\sf 1D}/J_{\sf 2D}=0.075$ similar to QMC. In both reference systems, we assumed $J_{\sf 1D}=J_{\sf 2D}=1$. The ordering temperature $T_N$ is the point, where $\bar\chi^s(T)$ reaches $(zJ_{i1}/J_{\sf 2D})^{-1}=50$. Then the reference quasi-2D system should order at $T_N/J_{\sf 2D}\simeq 0.35$ that compares quite well to the accurate QMC result $T_N/J_{\sf 2D}\simeq 0.326$ for the same interlayer coupling $J_{i1}/J_{\sf 2D}=0.01$.\cite{yasuda2005} We also get a reasonable agreement with QMC for the reference 1D system, where our mean-field theory (with $z=4$) predicts $T_N/J\simeq 0.029$, and the QMC result is $T_N/J\simeq 0.021$.\cite{yasuda2005} Finally, the LRO in CuP$_2$O$_6$ should appear below $T_N/J_{\sf 2D}\simeq 0.236$, which is again in good agreement, but now with the experimental value of $T_N/J_{\sf 2D}\simeq 0.2$.

More generally, our mean-field analysis elucidates the mechanism of the long-range ordering in CuP$_2$O$_6$ and in any system comprising both 1D and 2D sublattices. For a given interchain/interlayer coupling $J_{i1}$, the ordering temperature $T_N$ is determined by spin fluctuations in the 1D/2D units. The drastic difference between the ordering temperatures of quasi-1D and quasi-2D magnets is reflected by the large difference in their staggered susceptibilities. In a composite 2D+1D system, the relevant quantity is the geometric mean value of the staggered susceptibilities for 1D and 2D.

\section{Discussion and Summary}
CuP$_2$O$_6$ shows a rare example of coexisting spin sublattices with mixed dimensionalities and energy scales. Thermodynamic properties of this compound are well explained by a superposition of the 1D and 2D sublattices. However, a phenomenological analysis of the data is not straight-forward. For example, CuP$_2$O$_6$ does not show the broad maximum in the susceptibility, which would be typical for low-dimensional antiferromagnets. Nevertheless, this compound is clearly a low-dimensional antiferromagnet, as evidenced by our microscopic analysis and by the magnetic specific heat (Fig.~\ref{fig:heat}). Surprisingly, the decomposition of the overall 3D spin lattice into two low-dimensional sublattices remains valid at low temperatures, even below the N\'eel temperature $T_N$, where the LRO state involving both sublattices is formed.

The N\'eel temperature of $T_N/J_{\sf 2D}\simeq 0.2$ is remarkably low for a quasi-2D antiferromagnet. A simple mean-field treatment shows that the coupling of 2D units via 1D units or, more generally, via any spin units with strong spin fluctuations will reduce the ordering temperature dramatically. This system follows an ``averaged'' scenario, which is intermediate between 2D and 1D. More precisely, the ordering is described at the mean-field level by the geometrical mean of the staggered susceptibilities of the 1D and 2D spin  sublattices. 

Another interesting aspect is the ordered state of CuP$_2$O$_6$. The ordering in the 2D sublattice manifests itself by the weak net moment observed below $T_N$. Although this order has not been seen by neutron diffraction,\cite{gunsser1989} both ESR and the formation of net moment safely confirm the onset of the long-range magnetic order at 8~K, whereas the lack of the magnetic reflections may be ascribed to an insufficient sensitivity of the neutron experiment. Indeed, in a quasi-2D system, an ordered moment of only 0.6~$\mu_B$ is expected.\cite{sandvik1997} The persistent paramagnetic ESR signal below $T_N$ gives strong evidence for the very weak Cu2 moments along the chains. Further studies of the ordered state of CuP$_2$O$_6$ with neutron scattering and/or $^{31}$P nuclear magnetic resonance are highly desirable and are presently underway.

Surprisingly, only a few systems combining different types of spin lattices have been reported so far. The combination of spin chains and 2D square-lattice units is typical for the family of YBa$_2$Cu$_3$O$_{6+\delta}$ high-temperature superconductors, but the relevant charge-ordered state has never been achieved there. Indeed, at $\delta=0.5$ all Cu ions should be in the 2+ spin-$\frac12$ state, but residual charge transfer leads to a doping and superconductivity,\cite{jorgensen1990} so that the purely spin physics of a magnetic insulator can not be probed. We also note an interesting, albeit purely phenomenological similarity between CuP$_2$O$_6$ and CuB$_2$O$_4$, where three consecutive magnetic transitions have been observed.\cite{petrakovskii1999,roessli2001,*boehm2003,fukaya2006} Although several studies consider the spin system of CuB$_2$O$_4$ as a combination of square planes with strong couplings (compare to the Cu1 planes in CuP$_2$O$_6$) and interstitial Cu$^{+2}$ sites with weak couplings (compare to the Cu2 chains in CuP$_2$O$_6$),\cite{[{For example: }][]martynov2007} details of the microscopic magnetic model are far from being understood. Moreover, some of the magnetic structures are incommensurate and likely influenced by substantial DM couplings. The case of CuP$_2$O$_6$ should be more simple, because only one of the couplings features the DM component. This component is quite small and leads to a trivial spin canting without the complex helical modulation observed in CuB$_2$O$_4$.

In summary, we have studied thermodynamic properties and developed a microscopic magnetic model for CuP$_2$O$_6$. This compound features two weakly coupled spin sublattices with different dimensionalities and different energy scales. It undergoes a long-range ordering transition at $T_N\simeq 8$~K corresponding to $T_N/J_{\sf 2D}\simeq 0.2$, which is unusually low for a quasi-2D spin system. The DM couplings in the Cu1 sublattice induce a weak, but clearly visible spin canting. Moreover, the  paramagnetic ESR signal below 8~K puts forward substantial fluctuations that persist even below $T_N$. We argue that the coupling of 2D sublattices via 1D spin units hinders an overall 3D long-range magnetic order. Details of the long-range-ordered state in CuP$_2$O$_6$ deserve further investigation.

\begin{acknowledgments}
AT and IR acknowledge fruitful discussions with Oleg Janson. Part of this work has been supported by EuroMagNET II under the EU contract number 228043. RN was funded by MPG-DST (Max Planck Gesellschaft, Germany and Department of Science and Technology, India) fellowship. AT acknowledges financial support from the Mobilitas program of the ESF (grant MTT77). FA thanks GENCI (grant 2013-x2013050225) and CALMIP for the use of their numerical resources. IR acknowledges support from the Deutsche Forschungsgemeinschaft (DFG) under the Emmy-Noether program.
\end{acknowledgments}


\begin{thebibliography}{70}
\bibitem [{foo()}]{foot1}%
  \BibitemOpen
  \href@noop {} {}\bibinfo {note} {Here, we use the orthogonal $xyz$ coordinate
  frame defined as follows: $\yv\!=\!\bv$, $\zv\!=\!\cv$, and $\xv\!\perp\!
  \bv,\cv$. Owing to the monoclinic symmetry of CuP$_2$O$_6$, the $x$ direction
  deviates from the crystallographic $a$ direction by
  $28.58^{\circ}$.}\BibitemShut {Stop}%
\end{thebibliography}
%

\end{document}